\documentclass{llncs}
\usepackage{llncsdoc}
\usepackage{amsmath}
\usepackage{amssymb,bm}
\usepackage{color,xcolor,dsfont}
\usepackage{graphicx}
\usepackage[colorlinks=true, linkcolor=blue, urlcolor=blue, citecolor=blue, hyperindex, breaklinks]{hyperref}
\usepackage[matrix,frame,arrow]{xypic}
\usepackage[braket, qm]{qcircuit}
\usepackage{xspace}
\usepackage{algorithm}
\usepackage{algorithmic}
\usepackage{multicol}
\usepackage{multirow}
\usepackage{ulem}
\usepackage{float}

\usepackage{lineno}
\newcommand\add[1]{#1}

\newcommand\rev[2]{#2}

\newtheorem{assumption}{Assumption}
\newtheorem{metric}{Cost Metric}

\begin{document}

%\title{Security of SHA against a quantum adversary}
\title{Estimating the cost of generic quantum pre-image attacks on SHA-2 and SHA-3}

\author{Matthew Amy\inst{1,4} \and Olivia Di Matteo\inst{2,4} \and Vlad Gheorghiu\inst{3,4} \and Michele Mosca\inst{3,4,5,6} \and Alex Parent\inst{2,4} \and John Schanck\inst{3,4}}

\institute{David R. Cheriton School of Computer Science, University of Waterloo, Canada
\and Department of Physics \& Astronomy, University of Waterloo, Canada
\and Department of Combinatorics \& Optimization, University of Waterloo, Canada 
\and Institute for Quantum Computing, University of Waterloo, Canada 
\and Perimeter Institute for Theoretical Physics, Canada 
\and Canadian Institute for Advanced Research, Canada}

\maketitle

\begin{abstract}
We investigate the cost of Grover's quantum search algorithm when used in the
context of pre-image attacks on the SHA-2 and SHA-3 families of
hash functions.  Our cost model assumes that the attack is run on a surface
code based fault-tolerant quantum computer. Our estimates rely on a time-area
metric that costs the number of logical qubits times the depth of the circuit
in units of surface code cycles.  As a surface code cycle involves a
significant classical processing stage, our cost estimates allow for crude, but
direct, comparisons of classical and quantum algorithms.

We exhibit a circuit for a pre-image attack on SHA-256 that is
approximately $2^{153.8}$ surface code cycles deep and requires approximately
$2^{12.6}$ logical qubits. This yields an overall cost of $2^{166.4}$
logical-qubit-cycles.  Likewise we exhibit a SHA3-256 circuit that is
approximately $2^{146.5}$ surface code cycles deep and requires approximately
$2^{20}$ logical qubits for a total cost of, again, $2^{166.5}$
logical-qubit-cycles. Both attacks require on the order of $2^{128}$ queries in
a quantum black-box model, hence our results suggest that executing these
attacks may be as much as $275$ billion times more expensive than one would
expect from the simple query analysis.
\end{abstract}


%\pacs{03.67.Mn}

%%%%%%%%%%%%%%%%%%%%%%%%%%%%%%%%%%%%%%%%%%%%%%%%%%%%%%%%
%%%%%%%%%%%%%%%%%%%%%%%%%%%%%%%%%%%%%%%%%%%%%%%%%%%%%%%%
% Vlad & John
%%%%%%%%%%%%%%%%%%%%%%%%%%%%%%%%%%%%%%%%%%%%%%%%%%%%%%%%
%%%%%%%%%%%%%%%%%%%%%%%%%%%%%%%%%%%%%%%%%%%%%%%%%%%%%%%%

\setcounter{footnote}{0}
\section{Introduction\label{sct:intro}}
%\vlad{FIX the units (notation inconsistencies italic vs non-italic), fix the section labels}

%\vlad{FIX the fonts for the SHA primitives, in SHA3 we use something like ${\sf pad}10$, use the same for SHA2}

Two quantum algorithms threaten to dramatically reduce the security of
currently deployed cryptosystems: Shor's algorithm solves the abelian hidden
subgroup problem in polynomial time \cite{SJC.26.1484,Boneh1995}, and
Grover's algorithm provides a quadratic improvement in the number of queries
needed to solve black-box search problems 
\cite{PhysRevLett.79.325,PROP:PROP493,BBHT:2000}.

Efficient quantum algorithms for integer factorization, finite field discrete
logarithms, and elliptic curve discrete logarithms can all be constructed by
reduction to the abelian hidden subgroup problem. As such, cryptosystems based
on these problems can not be considered secure in a post-quantum environment.
Diffie-Hellman key exchange, RSA encryption, and RSA signatures will all need
to be replaced before quantum computers are available.  Some standards bodies
have already begun discussions about transitioning to new public key
cryptographic primitives \cite{NSA2015,NIST2016}.

The situation is less dire for hash functions and symmetric ciphers. In a
pre-quantum setting, a cryptographic primitive that relies on the hardness of
inverting a one-way function is said to offer $k$-bit security if
inverting the function is expected to take $N=2^k$ evaluations of the
function. An exhaustive search that is expected to take $O(N)$ queries with classical
hardware can be performed with $\Theta(\sqrt{N})$ queries using Grover's algorithm on
quantum hardware. Hence, Grover's algorithm could be said to reduce the
bit-security of such primitives by half; one might say that a 128-bit pre-quantum
primitive offers only 64-bit security in a post-quantum setting.

A conservative defense against quantum search is to double the security
parameter (e.g. the key length of a cipher, or the output length of a hash
function). However, this does not mean that the true cost of Grover's algorithm
should be ignored. A cryptanalyst may want to know the cost of an attack even
if it is clearly infeasible, and users of cryptosystems may want to know the
minimal security parameter that provides ``adequate protection'' in the sense
of \cite{Lenstra2004,Lenstra2001,Blaze1996}.

  In the context of pre-image search on a hash function, the cost of a
  pre-quantum attack is given as a number of invocations of the hash function.
  If one assumes that quantum queries have the same cost as classical queries,
  then the query model provides a reasonable comparison between quantum and
  classical search. However, realistic designs for large quantum computers
  call this assumption into question.

  The main difficulty is that the coherence time of physical qubits is finite.
  Noise in the physical system will eventually corrupt the state of any long
  computation. If the physical error rate can be suppressed below some
  threshold, then \emph{logical qubits} with arbitrarily long coherence times
  can be created using quantum error correcting codes. Preserving the state of
  a logical qubit is an active process that requires periodic evaluation of an
  error detection and correction routine. This is true even if no logical gates
  are performed on the logical qubit. Hence the classical processing required
  to evaluate a quantum circuit will grow in proportion to both the depth of 
  the circuit and the number of logical qubits on which it acts.

  We suggest that a cost model that facilitates direct comparisons of classical
  and quantum algorithms should take the classical computation required for
  quantum error correction into consideration. Clearly such estimates
  will be architecture dependent, and advances in quantum
  computing could invalidate architectural assumptions.

  To better understand the impact of costing quantum error correction,
  we present an estimate of the cost of pre-image attacks on SHA-2 and
  SHA-3 assuming a quantum architecture based on the surface code with a
  logical Clifford+$T$ gate set. We execute the following procedure for
  each hash function. First, we implement the function as a reversible
  circuit\footnote{Reversibility is necessary for the hash function to
  be useful as a subroutine in Grover search.} over the Clifford+$T$
  gate set. We use a quantum circuit optimization tool, ``$T$-par''
  \cite{6899791}, to minimize the circuit's $T$-count and
  $T$-depth\footnote{The logical $T$ gate is significantly more
  expensive than Clifford group gates on the surface code.}. With the
  optimized circuit in hand we estimate the additional overhead of fault
  tolerant computation. In particular, we estimate the size of the
  circuits needed to produce the ancillary states that are consumed by
  $T$-gates.

\begin{figure}[!htb]
\centering
  \begin{minipage}{.49\textwidth}
      \centering
      \vspace{0.6cm}
      \includegraphics[scale = 0.48]{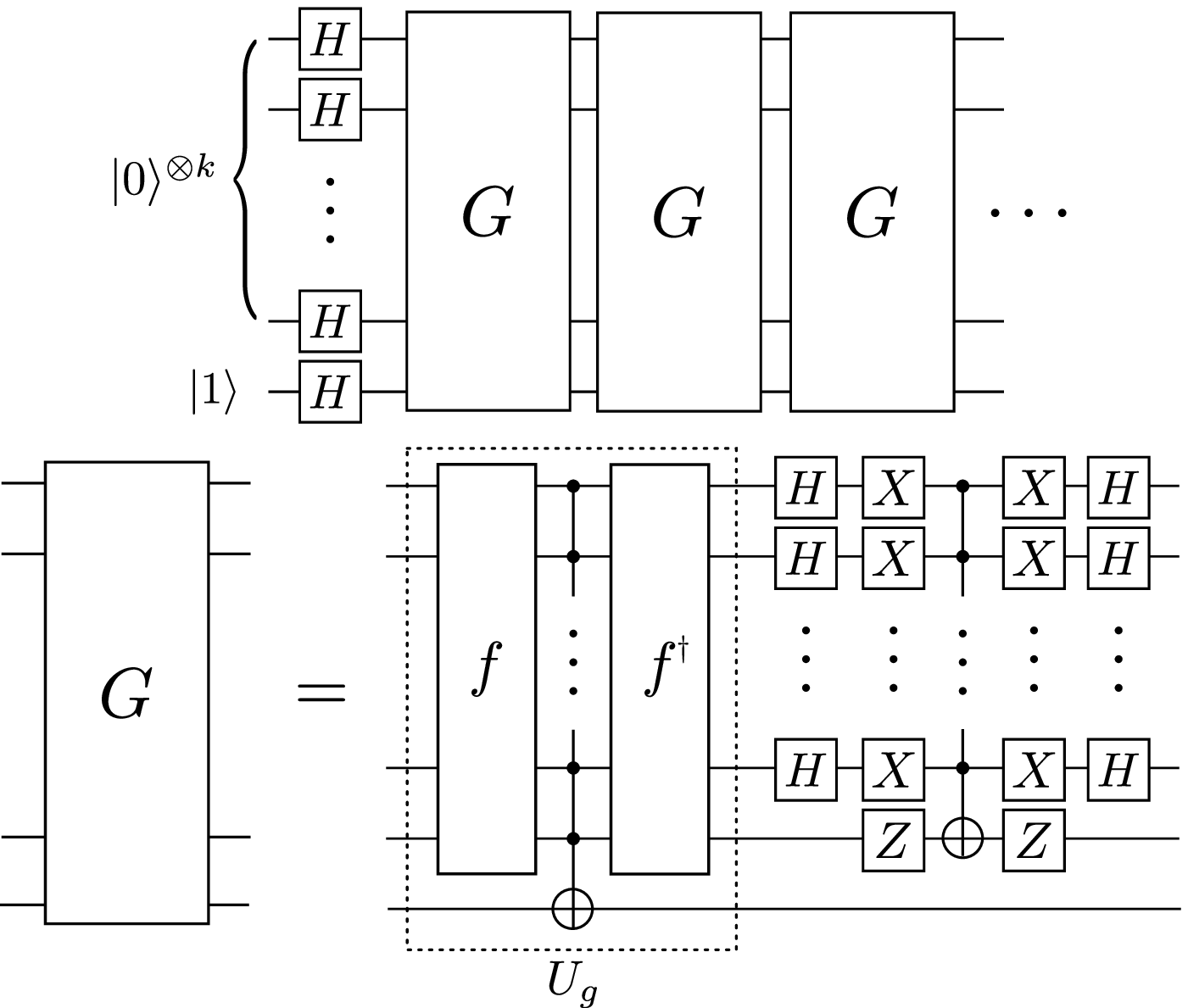}
      \vspace{-0.58cm}
      \caption{Grover searching with an oracle for $f : \{0,1\}^k \rightarrow \{0,1\}^k$.}
      \label{fig:full_algorithm}
  \end{minipage}
  \begin{minipage}{.49\textwidth}
  \centering
      \includegraphics[scale = 0.48]{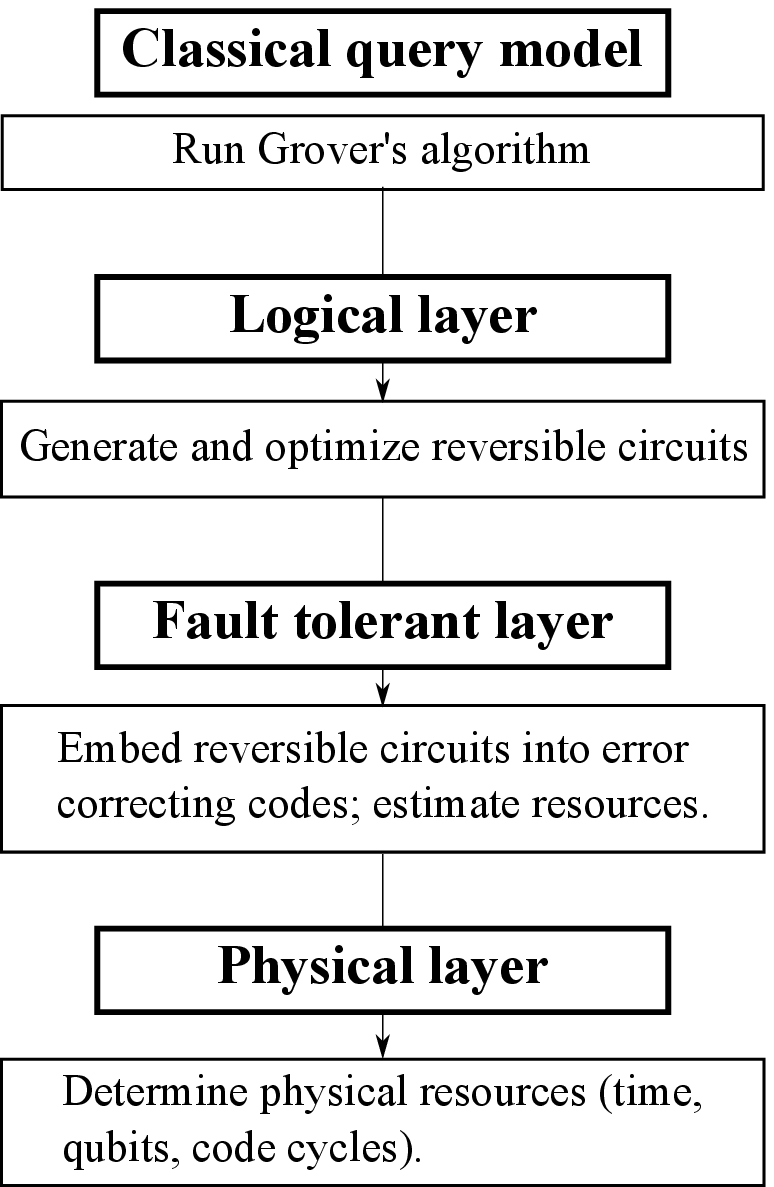}
      \caption{Analyzing Grover's algorithm.}
      \label{fig:flowchart_lite}
  \end{minipage}
\end{figure}

  Grassl et al. presented a logical-layer quantum circuit for applying
  Grover's algorithm to AES key recovery \cite{quantph.1512.04965}.
  Separately, Fowler et al.  have estimated the physical resources
  required to implement Shor's factoring algorithm on a surface code
  based quantum computer \cite{PhysRevA.86.032324}. Our resource
  estimates combine elements of both of these analyses. We focus on the
  number of logical qubits in the fault-tolerant circuit and the overall
  depth of the circuit in units of surface code cycles. While our cost
  model ties us to a particular quantum architecture, we segment our
  analysis into several layers so that the impact of a different
  assumptions at any particular level can be readily evaluated.
  We illustrate our method schematically in Fig.~\ref{fig:flowchart_lite}.
%  and in more detail in Fig.~\ref{fig:flowchart_full}.

  %Each surface code cycle
  %involves the execution of a classical syndrome decoding routine for every
  %logical qubit.  Thus in estimating these quantities we obtain the cost of a
  %pre-image attack purely in terms of classical computing resources.
  %Separately, we obtain an estimate for the number of physical qubits required
  %for the circuit, and an estimate for the wall-clock time of the computation.

\add{The structure of this article reflects our workflow. In Section~\ref{sct:grover} we state the problem of pre-image search using Grover's algorithm. Section \ref{sct:cost_metric} introduces our framework for computing costs, and Section \ref{sct2} applies these principles to compute the intrinsic cost of performing Grover search. Sections~\ref{sct:sha2} and \ref{sct:sha3} detail our procedure for generating reversible circuits for SHA-256 and SHA3-256 respectively. In Section \ref{sct:fault_tol} we embed these reversible implementations into a surface code, and estimate the required physical resources. We summarize our results and propose avenues of future research in Section \ref{sct:conclusions}.}

%%%%%%%%%%%%%%%%%%%%%%%%%%%%%%%%%%%%%%%%%%%%%%%%%%%%%%%%
%%%%%%%%%%%%%%%%%%%%%%%%%%%%%%%%%%%%%%%%%%%%%%%%%%%%%%%%
%%%%%%%%%%%%%%%%%%%%%%%%%%%%%%%%%%%%%%%%%%%%%%%%%%%%%%%%
%%%%%%%%%%%%%%%%%%%%%%%%%%%%%%%%%%%%%%%%%%%%%%%%%%%%%%%%
\section{Pre-image search via Grover's algorithm\label{sct:grover}}
%\john{Merge this with intro}
Let $f : \{0,1\}^k \rightarrow \{0,1\}^k$ be an efficiently function.
For a fixed $y \in \{0,1\}^k$, the value $x$ such that $f(x) = y$
is called a \emph{pre-image} of $y$. In the worst case, the only way to compute a pre-image of $y$ is to
systematically search the space of all inputs to $f$. A function that
must be searched in this way is known as a \emph{one-way function}. A
one-way function that is bijective is a
\emph{one-way permutation}\footnote{A
hash function that has been restricted to length $k$ inputs is expected
to behave roughly like a one-way permutation. The degree to which it
fails to be injective should not significantly affect the expected
probability of success for Grover's algorithm.}.

%\vlad{Mention that SHA is close to a one-way permutation}

Given a one-way permutation $f$, one might ask for the most cost
effective way of computing pre-images. With a classical computer one must query $f$
on the order of $2^k$ times before finding a pre-image. By contrast, a
quantum computer can perform the same search with $2^{k/2}$ queries to
$f$ by using Grover's algorithm \cite{PhysRevLett.79.325}. Of course, counting
only the queries to $f$ neglects the potentially significant overhead involved
in executing $f$ on a quantum computer.

Figure \ref{fig:full_algorithm} gives a high-level description of Grover's
algorithm. The algorithm makes $\lfloor \frac{\pi}{4} 2^{k/2}\rfloor$ calls to
$G$, the \emph{Grover iteration}. The Grover iteration has two
subroutines. The first, $U_g$, implements the predicate $g : \{0,1\}^k
\rightarrow \{0,1\}$ that maps $x$ to $1$ if and only if $f(x) =
y$. Each call to $U_g$ involves two calls to a reversible implementation of $f$
and one call to a comparison circuit that checks whether $f(x) = y$.

The second subroutine in $G$ implements the transformation $2 \op{0}{0} -
I$ and is called the \emph{diffusion operator}. The diffusion operator is
responsible for amplifying the probability that a measurement of the
output register would yield $x$ such that $f(x) = y$. As it involves
only single-qubit gates and a one $k$-fold controlled-NOT, the cost
of the diffusion operator is expected to be small compared with that of $U_g$.

%%%%%%%%%%%%%%%%%%%%%%%%%%%%%%%%%%%%%%%%%%%%%%%%%%%%%%%%
%%%%%%%%%%%%%%%%%%%%%%%%%%%%%%%%%%%%%%%%%%%%%%%%%%%%%%%%
%%%%%%%%%%%%%%%%%%%%%%%%%%%%%%%%%%%%%%%%%%%%%%%%%%%%%%%%
%%%%%%%%%%%%%%%%%%%%%%%%%%%%%%%%%%%%%%%%%%%%%%%%%%%%%%%%
\section{A cost metric for quantum computation\label{sct:cost_metric}}

%Commonly used cost metrics such as \emph{bit operations} \cites{} and
%\emph{dollar-days} \cites{} are difficult to apply to quantum computations.

% Bit operations are inadmissible simply because classical and quantum circuits
% employ distinct gate sets. One could attempt to compare counts of the primitive
% gates in each model, e.g. NAND vs Clifford+T gates, however this fails to be a
% fair comparison. Large quantum computations almost certainly cannot be
% performed without some form of error correction, and even the simplest
% Clifford+T circuits may require a significant number of physical gates per
% logical qubit operation.
%
% Comparing physical layer gates may provide a fair comparison, but even counting
% gates in the classical case is fraught with difficulties. How does one cost
% special purpose hardware, SIMD devices, or GPUs in terms of NANDs?

%Dollar-days -- the dollar cost of the machine times the duration of the
%computation -- are conceptually simple and allow non-technical users to perform
%risk calculations, but the economics of quantum computation are still largely
%uncertain.

\begin{quote}
  {\raggedleft \emph{Without significant future effort, the classical processing will
almost certainly limit the speed of any quantum computer, particularly
  one with intrinsically fast quantum gates.}}

  {\raggedleft Fowler--Whiteside--Hollenberg~\cite{Fowler2012a}\\}
\end{quote}

The majority of the overhead for quantum computation, under realistic
assumptions about quantum computing architectures, comes from error detection
and correction. There are a number of error correction methods in the literature, however the
most promising, from the perspective of experimental realizability, is the
surface code \cite{Fowler2012b}.

The surface code allows for the detection and correction of errors on a
two-dimensional array of nearest-neighbor coupled physical qubits.  A
distance $d$ surface code encodes a single logical qubit into an $n
\times n$ array of physical qubits $(n=2d-1)$. A classical error
detection algorithm must be run at regular intervals in order to track
the propagation of physical qubit errors and, ultimately, to prevent
logical errors. Every surface code \emph{cycle} involves some number of
one\add{-} and two\add{-}qubit physical quantum gates, physical qubit
measurements, and classical processing to detect and correct errors.

The need for classical processing allows us to make a partial comparison between
\add{the} cost of classical and quantum algorithms for any classical cost metric. The
fact that quantum system engineers consider classical processing
to be a bottleneck for quantum computation \cite{Fowler2012a} suggests that an analysis of the
classical processing may serve as a good proxy for an analysis of the cost of
quantum computation itself.

Performing this analysis requires that we make a number of assumptions about
how quantum computers will be built, not least of which is the assumption that
quantum computers will require error correcting codes, and that the surface
code will be the code of choice.

\begin{assumption}\label{assume:scqc}
  The resources required for any large quantum computation are well
  approximated by the resources required for that computation on a
  surface code based quantum computer.
\end{assumption}

Fowler et al.  \cite{Fowler2012} give an algorithm for the classical processing
required by the surface code. A timing analysis of this algorithm was given in
\cite{Fowler2012a}, and a parallel variant was presented in \cite{Fowler2013}.
Under a number of physically motivated assumptions, the algorithm of
\cite{Fowler2013} runs in constant time per round of error detection. It
assumes a quantum computer architecture consisting of an $L \times L$ grid of logical
qubits overlaid by a constant density mesh of classical computing units. More
specifically, the proposed design involves one ASIC (application-specific
integrated circuit) for each block of $C_a \times C_a$ physical qubits. These
ASICs are capable of nearest-neighbor communication, and the number of rounds
of communication between neighbors is bounded \add{with respect to} the error
model. The number of ASICs scales linearly with the number of logical qubits,
but the constant $C_a$, and the amount of computation each ASIC performs per
time step, is independent of the number of logical qubits.

Each logical qubit is a square grid of $n \times n$ physical qubits where $n$
depends on the length of the computation and the required level of error
suppression. We are able to estimate $n$ directly (Section
\ref{sct:fault_tol}). Following \cite{Fowler2012a} we will assume that $C_a =
n$. The number of classical computing units we estimate is therefore equal to
the number of logical qubits in the circuit.  Note that assuming $C_a = n$
introduces a dependence between $C_a$ and the length of the computation, but we
will ignore this detail. Since error correction must be performed on the time
scale of hundreds of nanoseconds ($200 ns$ in \cite{Fowler2012b}), we do not
expect it to be practical to make $C_a$ much larger than $n$. Furthermore,
while $n$ depends on the length of the computation it will always lie in a
fairly narrow range. A value of $n < 100$ is sufficient even for the extremely
long computations we consider. The comparatively short modular exponentiation
computations in \cite{Fowler2012b} require $n > 31$. As long as it is not
practical to take $C_a$ much larger than $100$, the assumption that $C_a = n$
will introduce only a small error in our analysis.

\begin{assumption}\label{assume:asics}
  The classical error correction routine for the surface code on an $L \times L$ grid of
  logical qubits requires an $L \times L$ mesh of classical processors (i.e. $C_a = n$).
\end{assumption}

The algorithm that each ASIC performs is non-trivial and estimating its exact
runtime depends on the physical qubit error model.  In \cite{Fowler2012a}
evidence was presented that the error correction algorithm requires $O(C_a^2)$
operations, on average, under a reasonable error model.  This work considered a
single qubit in isolation, and some additional overhead would be incurred by
communication between ASICs. A heuristic argument is given in \cite{Fowler2013}
that the communication overhead is also independent of $L$, i.e. that the
radius of communication for each processor depends on the noise model but not
on the number of logical qubits in the circuit.

\begin{assumption}
  Each ASIC performs a constant number of operations per surface code cycle.
\end{assumption}

Finally we (arbitrarily) peg the cost of a surface code cycle to the cost of a
hash function invocation. If we assume, as in \cite{Fowler2012b}, that a
surface code cycle time on the order of $100 ns$ is achievable, then we are
assuming that each logical qubit is equipped with an ASIC capable of performing
several million hashes per second. This would be on the very low end of what is
commercially available for Bitcoin mining today \cite{bitcoin}, however the
ASICs used for Bitcoin have very large circuit footprints. One could
alternatively justify this assumption by noting that typical hash functions
require $\approx 10$ cycles per byte on commercial desktop CPUs
\cite{EBACS}. This translates to approximately $\approx 1000$ cycles per hash
function invocation. Since commercial CPUs operate at around 4~GHz, this
again translates to a few million hashes per second.

\begin{assumption}\label{assume:equalcost}
  The temporal cost of one surface code cycle is equal to the temporal cost of one hash function invocation.
\end{assumption}

Combining Assumptions \ref{assume:scqc}, \ref{assume:asics}, and
\ref{assume:equalcost} we arrive at the following metric for comparing the
costs of classical and quantum computations.

\begin{metric}\label{metric}
  The cost of a quantum computation involving $\ell$ logical qubits for a duration of \add{$\sigma$} surface
  code cycles is equal to the cost of classically evaluating a hash function $\ell \cdot \add{\sigma}$ times. Equivalently we will say that \emph{one logical qubit cycle} is
  equivalent to \emph{one hash function invocation}.
\end{metric}

We will use the term ``cost'' to refer either to logical qubit cycles or to hash
function invocations.

%%%%%%%%%%%%%%%%%%%%%%%%%%%%%%%%%%%%%%%%%%%%%%%%%%%%%%%%
%%%%%%%%%%%%%%%%%%%%%%%%%%%%%%%%%%%%%%%%%%%%%%%%%%%%%%%%
% Vlad & John
%%%%%%%%%%%%%%%%%%%%%%%%%%%%%%%%%%%%%%%%%%%%%%%%%%%%%%%%
%%%%%%%%%%%%%%%%%%%%%%%%%%%%%%%%%%%%%%%%%%%%%%%%%%%%%%%%
\section{Intrinsic cost of Grover search\label{sct2}}

Suppose there is polynomial overhead per Grover iteration, i.e.
$\Theta(2^{k/2})$ Grover iterations cost $\approx k^v2^{k/2}$ logical qubit
cycles for some real $v$ independent of $k$. Then an adversary who is willing
to execute an algorithm of cost $2^C$ can use Grover's algorithm to search a
space of $k$ bits provided that
\begin{equation} \label{eq:fixedcostk}
  k/2 + v\log_2(k) \le C.
\end{equation}

We define the \emph{overhead} of the circuit as $v$ and the \emph{advantage} of
the circuit as $k/C$. Note that if we view $k$ as a function of $v$ and
$C$ then for any fixed $v$ we have
$\lim_{C\rightarrow \infty} k(v,C)/C = 2,$
i.e. asymptotically, Grover's algorithm provides a quadratic advantage over
classical search. However, here we are interested in non-asymptotic advantages.

When costing error correction, we must have $v\ge1$ purely from
the space required to represent the input. However, we should not
expect the temporal cost to be independent of $k$. Even if the temporal
cost is dominated by the $k$-fold controlled-NOT gate, the Clifford+$T$
depth of the circuit will be at least $\log_2(k)$ \cite{Selinger:13}.
Hence, $v \ge 1.375$ for $k\le256$.  This
still neglects some spatial overhead required for magic state
distillation, but $v = 1.375$ may be used to derive strict upper bounds,
in our cost model, for the advantage of Grover search.

% SHA2 depth 830720    qubits 2337     + 3615 ftqubits
% SHA3 depth  11040    qubits 3200     + 3615 * 13 ftqubits
% AES  depth 130929    qubits 1336

In practice the overhead will be much greater. The AES-256 circuit from
\cite{quantph.1512.04965} has depth $130929$ and requires $1336$ logical
qubits. This yields overhead of $v \approx 3.423$ from the reversible
layer alone.

Substituting $z = \frac{k\ln{2}}{2v}$, the case
of equality in Equation \ref{eq:fixedcostk} is
\begin{equation}\label{eq:kfromcv}
ze^z = \frac{2^{C/v}\ln{2}}{2v} \quad \Longrightarrow \quad
k(v,C) = \frac{2v}{\ln(2)} \cdot \operatorname{W}\left(\frac{2^{C/v} \ln 2}{2v}\right)
\end{equation}
where $\operatorname{W}$ is the Lambert W-function. Table
\ref{tab:kCtov} in Appendix \ref{app:tables} gives the advantage of
quantum search as a function of its cost $C$ and overhead $v$; $k$ is
computed using Equation \ref{eq:kfromcv}.

% Table \ref{tab:katoC} gives the cost of quantum pre-image search on a $k$-bit
% function. Table \ref{tab:crossover} gives the $k$ for which classical and quantum
% search have equal cost as a function of the overhead.

%%%%%%%%%%%%%%%%%%%%%%%%%%%%%%%%%%%%%%%%%%%%%%%%%%%%%%%%
%%%%%%%%%%%%%%%%%%%%%%%%%%%%%%%%%%%%%%%%%%%%%%%%%%%%%%%%
% Alex
%%%%%%%%%%%%%%%%%%%%%%%%%%%%%%%%%%%%%%%%%%%%%%%%%%%%%%%%
%%%%%%%%%%%%%%%%%%%%%%%%%%%%%%%%%%%%%%%%%%%%%%%%%%%%%%%%

\section{Reversible implementation of a SHA-256 oracle\label{sct:sha2}}

\rev{In this section we give a brief overview of the Secure Hash Algorithm 2 standard\cite{SHA2}. We then describe our implementation, using digest size 256, as a reversible circuit and report resource counts both at the reversible and Clifford+$T$ levels.}{}

%\subsection{Overview}

\rev{SHA-2}{The Secure Hash Algorithm $2$ (SHA-2) \cite{SHA2}} is a family of collision resistant cryptographic hash functions.
There are a total of six functions in the SHA-2 family: SHA-224, SHA-256, SHA-384, SHA-512, SHA-512/224 and SHA-512/256.
There are currently no known classical pre-image attacks against any of the SHA-2 algorithms which are faster then brute force.
We will focus on SHA-256, a commonly used variant\rev{.
For each we}{, and} will assume a message size of one block (512 bits).

First the message block is stretched using Algorithm~\ref{alg:sha2Stretch} and the result is stored in $\mathbf{W}$.
The internal state is then initialized using a set of constants.
The round function is then run 64 times,
each run \rev{uses}{using} a single entry of $\mathbf{W}$ to modify the internal state.
The round function for SHA-256 is shown in Algorithm~\ref{alg:sha2}.

\begin{algorithm}[H]
\caption{SHA-256. All variables are 32-bit words.}
\label{alg:sha2}
\begin{algorithmic}[1]
  \FOR{i=0 \TO 63}
%  \begingroup
%  \setlength{\itemindent}{2em}
%  \addtolength{\algorithmicindent}{\itemindent}
    \STATE \hspace*{1em} $\Sigma_1\gets (\mathbf{E} \ggg 6) \oplus (\mathbf{E} \ggg 11) \oplus  (\mathbf{E} \ggg 25)$
    \STATE \hspace*{1em} $\mathbf{Ch}\gets(\mathbf{E} \land \mathbf{F}) \oplus ( \neg\mathbf{E}\land \mathbf{G})$
    \STATE \hspace*{1em} $\text{t}_1\gets \mathbf{H} + \Sigma_1 + \mathbf{Ch} + \mathbf{K}[i] + \mathbf{W}[i]$
    \STATE \hspace*{1em} $\Sigma_0\gets(\mathbf{A} \ggg 2) \oplus (\mathbf{A} \ggg 13) \oplus (\mathbf{A} \ggg 22)$
    \STATE \hspace*{1em} $\text{Maj}\gets(\mathbf{A} \land \mathbf{B}) \oplus (\mathbf{A} \land \mathbf{C}) \oplus (\mathbf{B}\land\mathbf{C})$
    \STATE \hspace*{1em} $\mathbf{t}_2 \gets \Sigma_0 + \text{Maj}$
    \STATE \hspace*{1em} $\mathbf{H} \gets \mathbf{G}$
    \STATE \hspace*{1em} $\mathbf{G} \gets \mathbf{F}$
    \STATE \hspace*{1em} $\mathbf{F} \gets \mathbf{E}$
    \STATE \hspace*{1em} $\mathbf{E} \gets \mathbf{D} + \text{t}_1$
    \STATE \hspace*{1em} $\mathbf{D} \gets \mathbf{C}$
    \STATE \hspace*{1em} $\mathbf{C} \gets \mathbf{B}$
    \STATE \hspace*{1em} $\mathbf{B} \gets \mathbf{A}$
    \STATE \hspace*{1em} $\mathbf{A} \gets \text{t}_1 + \text{t}_2$
%  \endgroup
  \ENDFOR
\end{algorithmic}
\end{algorithm}

\begin{algorithm}[H]
\caption{SHA-256 Stretch. All variables are 32-bit words.}
\label{alg:sha2Stretch}
\begin{algorithmic}[1]
  \FOR{$i=16$ \TO 63}
%  \begingroup
%  \setlength{\itemindent}{2em}
%  \addtolength{\algorithmicindent}{\itemindent}
  	\STATE \hspace*{1em} $\sigma_0 \gets (\mathbf{W}_{i-15} \ggg 7) \oplus (\mathbf{W}_{i-15} \ggg 18) \oplus (\mathbf{W}_{i-15} \gg 3)$
  	\STATE \hspace*{1em} $\sigma_1 \gets (\mathbf{W}_{i-2} \ggg 17) \oplus (\mathbf{W}_{i-2} \ggg 19) \oplus (\mathbf{W}_{i-2} \gg 10)$
  	\STATE \hspace*{1em} $w[i] \gets  \mathbf{W}_{i-16} + \sigma_0 + \mathbf{W}_{i-7} + \sigma_1$
%  \endgroup
  \ENDFOR
\end{algorithmic}
\end{algorithm}

\begin{figure*}
  \centering
  \includegraphics[scale = 0.3]{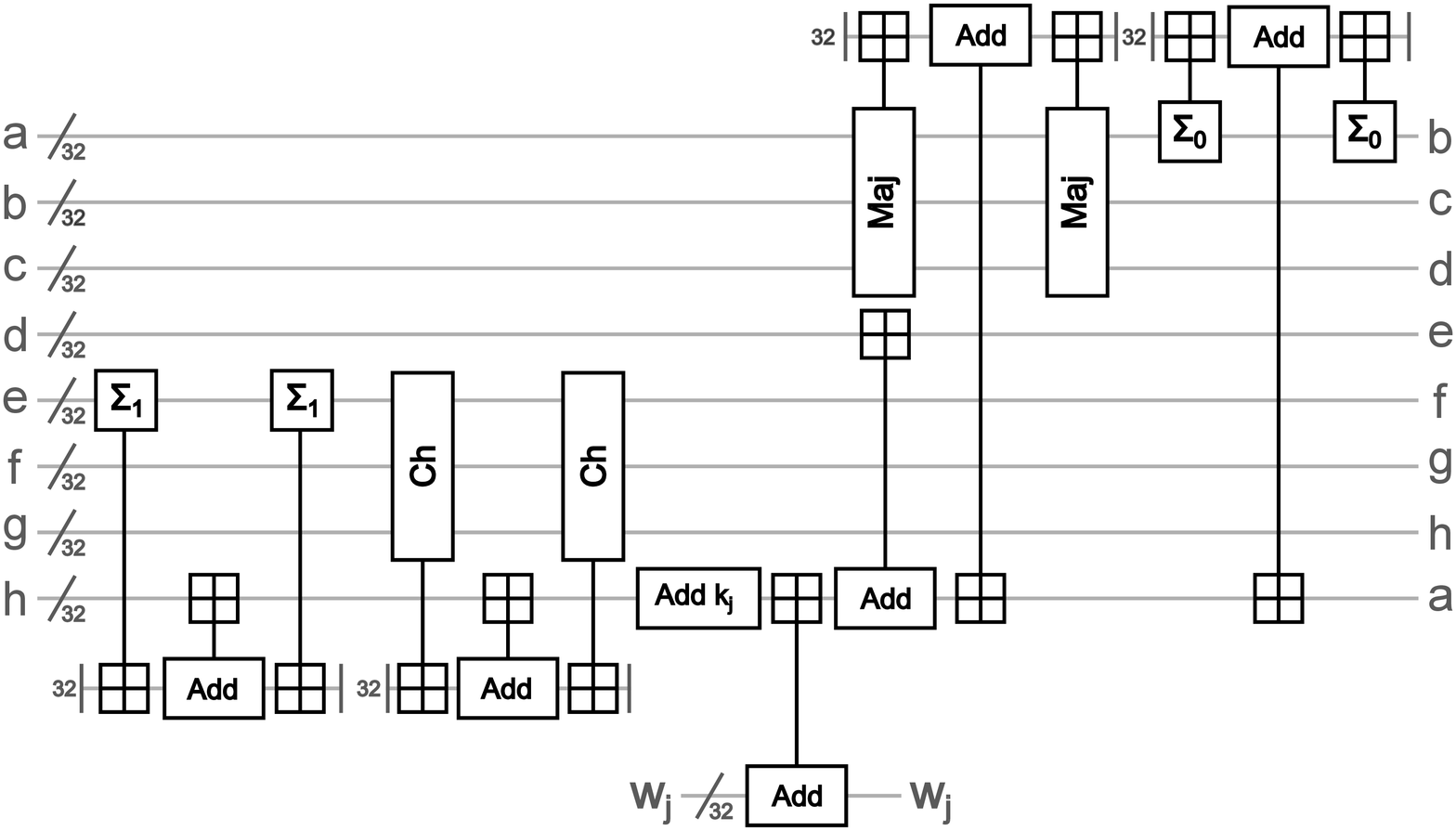}
  \caption{SHA-256 round.}
  \label{fig:sha2}
\end{figure*}

\subsection{Reversible implementation}

\rev{The circuit for SHA-256 construction}{Our implementation of
the SHA-256 algorithm as a reversible circuit} is similar to the one presented in
\cite{rev15} (with the addition of the stretching function).  Each round can be
performed fully reversibly (with access to the input) so no additional space is
accumulated as rounds are \add{performed}. The in-place adders shown in the circuit
are described in \cite{cuccaro04}. The adders perform the function $(a,b,0)
\mapsto (a,a+b,0)$ where the $0$ is a single ancilla bit used by the adder.
Since the $\Sigma$ blocks use only rotate and XOR operations, they \rev{can be}{are}
constructed using CNOT gates exclusively.

$\sf{Maj}$ is the bitwise majority function.  The majority function \rev{can be}{is} computed
using \rev{using}{} a CNOT gate and two Toffoli gates as show in Fig.~\ref{fig:maj}.

\begin{figure}
\centering
\begin{minipage}{.49\textwidth}
    \centering
    \[
    \Qcircuit @C=1em @R=.7em {
        \lstick{a}  & \ctrl{1}  & \ctrl{1} & \qw      & \rstick{a}\qw \\
        \lstick{b}  & \ctrl{2}  & \targ    & \ctrl{1} & \rstick{a\oplus b}\qw \\
        \lstick{c}  & \qw       & \qw      & \ctrl{1} & \rstick{c}\qw \\
        \lstick{0}  & \targ     & \qw      & \targ    & \rstick{ab\oplus ac \oplus bc}\qw
    }
    \]
    \caption{Majority circuit implementation. The $a\oplus b$ line will be returned
             to $b$ when the inverse circuit is applied .}
    \label{fig:maj}
\end{minipage}
\begin{minipage}{.49\textwidth}
    \centering
    \[
    \Qcircuit @C=1em @R=.7em {
        \lstick{a} & \qw      & \qw       & \ctrl{3} & \rstick{a}\qw \\
        \lstick{b} & \qw      & \ctrl{1}  & \qw      & \rstick{b}\qw \\
        \lstick{c} & \ctrl{1} & \targ     & \ctrl{1} & \rstick{b \oplus c}\qw \\
        \lstick{0} & \targ    & \qw       & \targ    & \rstick{ab \oplus \neg ac}\qw
    }
    \]
    \caption{$\sf{Ch}$ circuit implementation.
             This circuit is applied bitwise to input of each $\sf{Ch}$ block.}
    \label{fig:ch}
\end{minipage}    
\end{figure}

The $\sf{Ch}$ function is $ab\oplus\neg ac$ which can be rewritten as $a(b\oplus c)
\oplus c$. This requires a single Toffoli gate as shown in Fig.~\ref{fig:ch}.
    
There are a few options for constructing the round circuit.  For example if
space is available some of the additions can be performed in parallel, and the
cleanup of the $\Sigma$, $\sf{Ch}$, and $\sf{Maj}$ functions can be neglected if it is desirable
to exchange space for a lower gate count. We select the round implementation
shown in Fig.~\ref{fig:sha2}.

\subsection{Quantum implementation}

For the quantum implementation we \rev{must convert}{converted} the
Toffoli-CNOT-NOT circuit \rev{(refer explicitly to a
figre/equation)}{(Fig.~\ref{fig:sha2})} discussed above into a Clifford+$T$
circuit. To expand the Toffoli gates we used the $T$-depth $3$ Toffoli reported
in \cite{Amy:13}.  $T$-par was then used to optimize a single round. The results
are shown in table \ref{table:shaopt}.  Due to the construction of the adders
every Toffoli gate shares two controls with another Toffoli gate. This allows
$T$-par to remove a large number of $T$-gates (see \cite{Selinger:13}).

\begin{table}
\footnotesize
\centering
\begin{tabular}{lrrrrrrrr}
\hline \hline
                     & $T$/$T^\dagger$ & $P$/$P^\dagger$ & $Z$    &  $H$    & CNOT    & $T$-Depth & Depth \\ \hline
Round                & 5278          & 0             & 0    & 1508   & 6800    & 2262      & 8262  \\
Round (Opt.)         & 3020          & 931           & 96   & 1192   & 63501   & 1100      & 12980 \\
Stretch              & 1329          & 0             & 0    & 372    & 2064    & 558       & 2331  \\
Stretch (Opt.)       & 744           & 279           & 0    & 372    & 3021    & 372       & 2907 \\
\hline
SHA-256             & 401584        &     0         &    0 & 114368 &  534272 & 171552    & 528768 \\
SHA-256 (Opt.) & 228992        & 72976         & 6144 &  94144 & 4209072 & 70400     & 830720 \\ \hline \hline \\

\end{tabular}
\caption{
$T$-par optimization results for a single round of SHA-256, one iteration of the stretch algorithm and full SHA-256.
Note that 64 iterations of the round function and 48 iterations of the stretch function are needed.
The stretch function does not contribute to overall depth since it can be performed in parallel with the rounds function.
No $X$ gates are used so an $X$ column is not included.
The circuit uses 2402 total logical qubits.
}
\label{table:shaopt}
\end{table}

Observing that the depth of the optimized SHA-256 circuit, $830720$, is approximately
$256^{2.458}$, and likewise that it requires $2402 \approx 256^{1.404}$ logical qubits,
the overhead, from the reversible layer alone, is $v \approx 3.862$.

%%%%%%%%%%%%%%%%%%%%%%%%%%%%%%%%%%%%%%%%%%%%%%%%%%%%%%%%
%%%%%%%%%%%%%%%%%%%%%%%%%%%%%%%%%%%%%%%%%%%%%%%%%%%%%%%%
% Matt
%%%%%%%%%%%%%%%%%%%%%%%%%%%%%%%%%%%%%%%%%%%%%%%%%%%%%%%%
%%%%%%%%%%%%%%%%%%%%%%%%%%%%%%%%%%%%%%%%%%%%%%%%%%%%%%%%

\newcommand{\keccak}{{\sc Keccak}\xspace}
\newcommand{\Z}{\mathbb{Z}}

\section{Reversible implementation of a SHA3-256 oracle\label{sct:sha3}}

\rev{In this section we provide a brief overview of and report resource counts for Secure Hash Algorithm 3.}{}

% \subsection{Overview}

The Secure Hash Algorithm 3 standard \cite{SHA3} defines six individual hash algorithms, based on the length of their output in the case of SHA3-224, SHA3-256, SHA3-384 and SHA3-512, or their security strength in the case of SHAKE-128 and SHAKE-256. In contrast to the SHA-2 standard, each of the SHA-3 algorithms requires effectively the same resources to implement reversibly, owing to their definition as \emph{cryptographic sponge functions} \cite{Sponge}. Analogous to a sponge, the \rev{general}{} sponge construction first pads the \rev{variable-length }{}input to a multiple of the given rate constant then absorbs the padded message in chunks, applying a permutation to the state after each chunk, before ``squeezing'' out a hash value of desired length. Each of the SHA-3 algorithms use the same underlying permutation, but vary the chunk size, padding and output lengths.

The full SHA-3 algorithm is given in pseudocode in Algorithm~\ref{alg:sha3}. Each instance\rev{, SHA3-$k$,}{} results from the sponge construction with permutation \keccak-$p[1600, 24]$ described below, padding function ${\sf pad}10^*1(x, m)$ which produces a length $-m\mod x$ string of the form (as a regular expression) $10^*1$, and rate $1600-2\add{k}$. \rev{In particular, t}{T}he algorithm first pads the input message $M$ with the string $0110^*1$ \rev{where the number of repetitions is chosen to make the total length a multiple of $1600-2\add{k}$}{to a total length some multiple of $1600-2\add{k}$}. It then splits up this string into length $1600-2\add{k}$ segments and absorbs each of these segments into the current hash value $S$ then applies the \keccak-$p[1600, 24]$ permutation. \rev{In the end it truncates the hash value}{Finally the hash value is truncated} to a length $\add{k}$ string.

The SHAKE algorithms are obtained by padding the input $M$ with a string of the form $111110^*1$, but otherwise proceed identically to SHA-3

\begin{algorithm}[H]
\caption{SHA3-$k(M)$.}
\label{alg:sha3}
\begin{algorithmic}[1]
\STATE $ P \gets M01({\sf pad}10^*1(1600 - 2\add{k}, |M|)$
\STATE Divide $P$ into length $1600-2\add{k}$ strings $P_1, P_2,\dots, P_n$
\STATE $S \gets 0^{1600}$
\FOR{i=1 \TO n}
%\begingroup
%\setlength{\itemindent}{2em}
%\addtolength{\algorithmicindent}{\itemindent}
	\STATE \hspace*{1em} $S \gets $\keccak-$p[1600, 24](S\oplus(P_i0^{2\add{k}}))$
%\endgroup
\ENDFOR
%\STATE $Z \gets \epsilon$
%\WHILE{|Z|<c}
%	\STATE $Z \gets ZS[0, 1600-2c]$
%	\STATE $S \gets \keccak-p[1600, 24](S)$
%\ENDWHILE
%\RETURN Z[0, c]
\RETURN $S[0, c-1]$
\end{algorithmic}
\end{algorithm}

\rev{}{Assuming the pre-image has length $k$,} the padded message $P$ has length exactly $1600 - 2\add{k}$ and hence $n=1$ for every value of $\add{k}$, so Algorithm~\ref{alg:sha3} \rev{effectively }{}reduces to one application of \keccak-$p[1600, 24]$.

\subsubsection{The \keccak permutation}

The permutation underlying the sponge construction in each SHA-3 variant is an instance of a family of functions, denoted \keccak-$p[b, \add{r}]$. The \keccak permutation accepts a 5 by 5 array of \emph{lanes}, bitstrings of length $w=2^l$ for some $l$ where $b=25w$, and performs $\add{r}$ rounds of an invertible operation on this array. In particular, round $i$ is defined, for $12+2l - \add{r}$ up to $12+2l - 1$, as $R_i=\iota_i\circ\chi\circ\pi\circ\rho\circ\theta$, where the component functions are described in \rev{a declarative fashion in }{}Figure~\ref{fig:rnd}. Note that array indices are taken mod $5$ and $A, A'$ denote the input and output arrays, respectively. \rev{Additionally, t}{T}he rotation array \rev{$r$}{$c$} and round constants $RC(i)$ are \rev{values which may be precomputed -- we do not give them here but refer the interested reader to \cite{Keccak}.}{pre-computed values.}

The \keccak-$p[b, \add{r}]$ permutation itself \rev{may then be}{is} defined as the composition of all $\add{r}$ rounds, indexed from $12+2l -\add{r}$ to $12+2l - 1$.%\keccak-$p[b,n_r]$ performs $n_r$ rounds applies the mapping $\iota(i)\circ\chi\circ\pi\circ\rho\circ\theta$
While any parameters could potentially be used to define a hash function, only \keccak-$p[1600, 24]$ is used in the SHA-3 standard. Note that the lane size $w$ in this case is 64 bits.

\begin{figure*}
\begin{align}
\theta&: A'[x][y][z] &\leftarrow& A[x][y][z] \oplus \left(\bigoplus_{y'\in\Z_4} A[x-1][y'][z] \oplus A[x+1][y'][z-1]\right) \label{eqn1} \\
\rho&: A'[x][y][z] &\leftarrow& A[x][y][z + \rev{r}{c}(x, y)] \label{eqn2} \\
\pi&: A'[y][2x+3y][z] &\leftarrow& A[x][y][z] \label{eqn3} \\
\chi&: A'[x][y][z] &\leftarrow& A[x][y][z] \oplus A[x+2][y][z]\oplus A[x+1][y][z])A[x+2][y][z] \label{eqn4} \\
\iota_i&: A'[x][y][z] &\leftarrow& A[x][y][z] \oplus RC(i)[x][y][z] \label{eqn5}
\end{align}
\caption{The component functions of $R_i$}\label{fig:comp}
\end{figure*}

\subsection{Reversible implementation}

%We describe our implementation of the 4 main SHA-3 functions. In each case, the padded message $P=M1010^*1$ is divided into $n$ length $1600 - 2c$ chunks, then each chunk is repeatedly XOR-ed into the hash register and \keccak-$p[1600, 24]$ is run (see Figure~\ref{fig:sha3}). The resulting circuit comprises $|P|$ CNOT gates, along with $n$ iterations of \keccak-$p[1600, 24]$. We describe our construction of the \keccak-$p$ circuit in detail and give precise resource counts for $b=1600, n_r=24$.

%\subsubsection{\keccak-$p[b, n_r]$}

\rev{We now describe our implementation of the \keccak family of permutation functions as reversible circuits.}{}
Given the large size of the input register for the instance used in SHA3-256 ($1600$ bits), we sought a space-efficient implementation as opposed to a more straightforward implementation using Bennett's method \cite{Bennett:73} which would add an extra $1600$ bits \emph{per round}, to a total of $38400$ bits. While this space usage could be reduced by using \emph{pebble games} \cite{Bennett:89}, the number of iterations of \keccak-$p$ would drastically increase. Instead, we chose to perform each round \emph{in place} by utilizing the fact that each component function $(\theta, \rho, \pi, \chi, \iota_i)$ is invertible. \rev{Specifically, we use their inverses to clear the inputs and reclaim them as ancillas.}{} The resulting circuit requires only a single temporary register the size of the input, which is returned to the all-zero state at the end of each round.

 %In particular, as each component function $(\theta, \rho, \pi, \chi, \iota)$ is invertible, we use their inverses rather than copy out the result and ``uncompute'' as in Bennett's method. The advantage with this method is that we use only a single temporary register the size of the input, whereas $\theta$ and $\chi$ both appear to require local temporary storage.

 %to perform each round in place, using only a temporary register the size of the input which is returned clean. By contrast, Bennett's method would require additional ancilla bits on top of the additional temporary register, as there appears to be no straightforward way to implement $\theta$ and $\chi$ without ancillas.

 \begin{figure*}
\[
\Qcircuit @C=1.13em @R=2em {
\lstick{A} & \qw & {/}\qw & \ustick{25w}  \qw & \multigate{1}{\theta} & \ustick{A}\qw & \multigate{1}{\theta^{-1}} & \qw {|} & & {|} & \multigate{1}{\chi} & \qw & \qw & \ustick{\chi\circ\pi\circ\rho\circ\theta(A)}\qw & \qw & \qw & \multigate{1}{\chi^{-1}} & \qw & \gate{\iota_i} & \rstick{R_i(A)} \qw  \\
 & {|} & {/}\qw &  \ustick{25w} \qw & \ghost{\theta} & \ustick{\theta(A)}\qw & \ghost{\theta^{-1}} & \qw & \gate{\pi\circ\rho} & \qw & \ghost{\chi} & \qw & \qw & \ustick{\pi\circ\rho\circ\theta(A)}\qw & \qw & \qw & \ghost{\chi^{-1}} & \qw {|} & & &
}
\]
\caption{Reversible circuit implementation for round $i$ of \keccak-$p$.}
\label{fig:rnd}
\end{figure*}
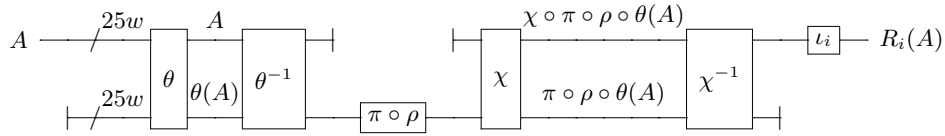

Fig.~\ref{fig:rnd} shows our circuit layout for a given round of \keccak-$p[b, \add{r}]$. We compute $\theta(A)$ into the ancilla register by a straightforward implementation of (\ref{eqn1}), as binary addition ($\oplus$) is implemented reversibly by the CNOT gate. The implementation of $\theta^{-1}:|\psi\rangle|\theta(A)\rangle\mapsto|\psi\oplus A\rangle|\theta(A)$ is much less obvious -- we adapted our implementation from the C++ library \keccak tools \cite{KeccakTools} with minor modifications to remove temporary registers. To reduce the number of unnecessary gates, we perform the $\rho$ and $\pi$ operations ``in software'' rather than physically swapping bits. The $\chi$ and $\chi^{-1}$ operations are again straightforward implementations of (\ref{eqn4}) and the inverse operation from \keccak tools, respectively, using Toffoli gates to implement the binary multiplications. Finally the addition of the round constant ($\iota_i$) is a sequence of at most $5$ NOT gates, precomputed for each of the 24 individual rounds. 

%Table~\ref{tab1} gives the ancilla, gate counts and Toffoli depth for our implementations of the component functions of \keccak-$p[1600, 24]$, as well as the total logical resources for the entire circuit. The $\rho$ and $\pi$ components are left out as they are implemented purely in software and don't require actual gates.

\rev{Table~\ref{tab3} gives the Clifford+$T$ resource counts for our implementations of the component functions of \keccak-$p[1600, 24]$, as well as the total logical resources for the entire circuit. The $\rho$ and $\pi$ components are left out as they are implemented purely in software and don't require actual gates.}{} As a function of the lane width $w$, $\theta$ comprises $275w$ \rev{bit additions, implemented with }{}CNOT gates. The inverse of $\theta$ is more difficult to assign a \rev{simple}{} formula to, as it depends on some precomputed constants -- in particular, $\theta{-1}$ is implemented using $125w\cdot \add{j}$ CNOT gates, where $\add{j}$ is 170 for $b=1600$. As $\rho$ and $\pi$ are implemented simply by re-indexing, they have no logical cost. We implement $\chi$ using $50w$ additions and $25w$ multiplications, giving $50w$ CNOT gates and $25w$ Toffoli gates in $5$ parallel stages. Finally $\chi^{-1}$ requires \rev{first }{}$25w$ CNOT gates to copy the output back into the initial register, then $60w$ CNOT and $30w$ Toffoli gates in $6$ parallel stages. As the cost of $\iota_i$ is dependent on the round, we don't give its per-round resources.

The final circuit comprises $3200$ qubits, $85$ NOT gates, $33269760$ CNOT gates and $84480$ Toffoli gates. Additionally, the Toffoli gates are arranged in $264$ parallel stages.

\subsection{Quantum implementation}

As with the Clifford+$T$ implementation of SHA-256, we used the
$T$-depth $3$ Toffoli reported in \cite{Amy:13} to expand each Toffoli
gate. Since the $\chi$ (and $\chi^{-1}$) transformation is the only
non-linear operation of \keccak-$p[1600, 24]$, we applied $T$-par just
to the $\chi/\chi^{-1}$ subcircuit to optimize $T$-count and depth. We
used the formally verified reversible circuit compiler {\sc ReVerC}
\cite{ReVerC} to generate a machine readable initial circuit for
$\chi/\chi^{-1}$ -- while the {\sc ReVerC} compiler performs some space
optimization \cite{ark16}, the straightforward manner in which we
implemented the circuit meant the compiled circuit coincided exactly
with our analysis above. The optimized results are reported in
Table~\ref{tab3}. Note that each algorithm in the SHA-3 family
corresponds to one application of \keccak-$p[1600, 24]$ for our
purposes, so the resources are identical for any output size.

%circuit was too large for $T$-par to handle, we individually optimized the $\chi$ and $\chi^{-1}$ subcircuit, which is the only non-linear operation (and hence, the only segment utilizing $T$ gates). The resulting circuit for $\chi$ and $\chi^{-1}$ contains $587520$ $T$-gates, a reduction of $3840$, with a $T$-depth of $18$, down from $33$. As expected there was a non-trivial increase in the number of CNOT gates -- from $???$ to $???$ -- but the gains in $T$-depth make up for these extra CNOT gates in the fault-tolerance resource estimate (see Section~\ref{sec:???}).

\begin{table}
\footnotesize
\centering
\begin{tabular}{l r r r r r r r}
\hline \hline
& $X$ & $P$/$P^\dagger$ & $T$/$T^\dagger$ & $H$ & CNOT & $T$-depth & Depth \\ \hline
$\theta$ & 0 & 0 & 0 & 0 & 17600 & 0 & 275 \\
$\theta^{-1}$ & 0 & 0 & 0 & 0 & 1360000 & 0 & 25 \\
$\chi$ & 0 & 0 & 11200 & 3200 & 14400 & 15 & 55 \\
$\chi^{-1}$ & 0 & 0 & 13440 & 3840 & 18880 & 18 & 66 \\
$\iota$ & 85 & 0 & 0 & 0 & 0 & 0 & 24 \\ \hline
{SHA3-256} & 85 & 0 & 591360 & 168960 & 33269760 & 792 & 10128  \\
{SHA3-256} (Opt.) & 85 & 46080 & 499200 & 168960 & 34260480 & 432 & 11040  \\ \hline \hline
\end{tabular}
\caption{Clifford+$T$ resource counts for the \keccak-$p[1600, 24]$ components, as well as for the full oracle implementation of SHA3-256. $\iota$ gives the combined resource counts for all $24$ rounds of $\iota_i$. The circuit uses 3200 total logical qubits.}
\label{tab3}
\end{table}

\rev{}{As an illustration of the overhead for SHA3-256, our reversible
SHA3-256 circuit, having depth $11040\approx256^{1.679}$ and a logical
qubit count of $3200\approx 256^{1.455}$ yields $v \approx 3.134$ at the reversible layer.}

%%%%%%%%%%%%%%%%%%%%%%%%%%%%%%%%%%%%%%%%%%%%%%%%%%%%%%%%
%%%%%%%%%%%%%%%%%%%%%%%%%%%%%%%%%%%%%%%%%%%%%%%%%%%%%%%%
% Vlad and Olivia
%%%%%%%%%%%%%%%%%%%%%%%%%%%%%%%%%%%%%%%%%%%%%%%%%%%%%%%%
%%%%%%%%%%%%%%%%%%%%%%%%%%%%%%%%%%%%%%%%%%%%%%%%%%%%%%%%
\section{Fault-tolerant cost\label{sct:fault_tol}}

The $T$ gate is the most expensive in terms of \rev{}{the} resources
needed for implementing a circuit fault-tolerantly in a surface code.
Most known schemes implement the $T$ gate using an auxiliary resource
called a \emph{magic state}. The latter is usually prepared in a faulty
manner, and purified to the desired fidelity via a procedure called
\emph{magic state distillation}. Fault-tolerant magic state
\emph{distilleries} (circuits for performing magic state distillation)
require a substantial number of logical qubits. In this section we
estimate the additional resources required by distilleries in the
particular case of SHA-256 and SHA3-256.

Let \add{$T^{c}_U$} denote the $T$-count of a circuit \add{$U$} (i.e., total number of logical $T$ gates), and let \add{$T^{d}_U$} be the $T$-depth of the circuit. We denote by \add{$T^{w}_U = T^c_U/T^d_U$} the $T$-width of the circuit (i.e., the number of logical $T$ gates that can be done in parallel on average for each layer of depth). Each $T$ gate requires one logical magic state of the form
\begin{equation}\label{eqn_ft_1}
\ket{A_L}:=\frac{\ket{0_L} + \mathrm{e}^{\mathrm{i}\pi/4}\ket{1_L}}{\sqrt{2}}
\end{equation}
for its implementation. For the \add{entirety of $U$} to run successfully, the magic states $\ket{A_L}$ have to be produced with an error rate no larger than
$p_{out} = 1/\add{T^{c}_U}$.

The magic state distillation procedure is based on the following scheme.
The procedure starts with a physical magic state prepared with some
failure probability $p_{in}$. This faulty state is then \emph{injected}
into an error correcting code, and then by performing a suitable
distillation procedure on the output carrier qubits of the encoded state
a magic state with a smaller failure probability is distilled. If this
failure probability is still larger than the desired $p_{out}$, the
scheme uses another layer of distillation, i.e. concatenates the first
layer of distillation with a second layer of distillation, and so forth.
The failure probability thus decreases exponentially.

In our case, we use the Reed-Muller 15-to-1 distillation scheme
introduced in \cite{PhysRevA.71.022316}. Given a state injection error
rate $p_{in}$, the output error rate after a layer of distillation can
be made arbitrarily close to the ideal $p_{dist}=35p_{in}^3$ provided we
ignore the logical errors that may appear during the distillation
procedure (those can be ignored if the distillation code uses logical
qubits with high enough distance). As pointed out in
\cite{Fowler:2013aa} logical errors do not need to be fully eliminated.
We also assume that the physical error rate per gate in the surface
code, $p_g$, is approximately 10 times smaller than $p_{in}$, i.e.
$p_g = p_{in}/10$,
as during the state injection approximately 10 gates have to perform
without a fault before error protection is available (see
\cite{PhysRevA.86.032324} for more details).

We define $\varepsilon$ so that $\varepsilon p_{dist}$ represents the
amount of logical error introduced, so $p_{out} =
(1+\varepsilon)p_{dist}$. In the balanced case $\varepsilon = 1$ the
logical circuit introduces the same amount of errors as distillation
eliminates. Algorithm~\ref{alg:statedistillation} \cite{Fowler:2013aa}
summarizes the procedure for estimating the number of rounds of state
distillation needed to achieve a given output error rate, as well as the
required minimum code distances at each round. Note that $d_1$
represents the distance of the surface code used in the top layer of
distillation (where by top we mean the initial copy of the Reed-Muller
circuit), $d_2$ the distance of the surface code used in the next layer,
and so forth.

\begin{algorithm}
\caption{Estimating the required number of rounds of magic state distillation and the corresponding distances of the concatenated codes}
\begin{algorithmic}[1]
  \STATE \textbf{Input:} $\varepsilon, p_{in}, p_{out}, p_g (=p_{in}/10)$
  \STATE $ d \leftarrow \text{empty list }[]$
  \STATE $p \leftarrow p_{out}$
  \STATE $i \leftarrow 0$
  \REPEAT

       \STATE \hspace*{1em} $i \leftarrow i + 1$
       \STATE \hspace*{1em} $p_i \leftarrow p$
       \STATE \hspace*{1em} Find minimum $d_i$ such that  $192 d_i (100 p_g)^{\frac{d_i+1}{2}} < \frac{\varepsilon p_i}{1 + \varepsilon} $
       \STATE \hspace*{1em} $p \leftarrow \sqrt[3]{ p_i / (35(1+\varepsilon))}$
       \STATE \hspace*{1em} $d$.append($d_i$)

  \UNTIL { $p  > p_{in}$ }
  \STATE \textbf{Output:} $d = [d_1, \ldots, d_i]$
\end{algorithmic}
\label{alg:statedistillation}
\end{algorithm}

%A summary of notation for the physical quantities we reference below is included in Fig. \ref{fig:flowchart_full}.

\subsection{SHA-256}\label{sct5A}

The $T$-count of our SHA-256 circuit is \add{$T^{c}_{\hbox{{\tiny
SHA-256}}} = 228992$} (see Table~\ref{table:shaopt}), and the $T$-count
of the $k$-fold controlled-NOT is \add{$T^{c}_{\hbox{{\tiny CNOT-k}}} =
32k-84$} \cite{quantph.1512.04965}. With $k = 256$, we have
$T^{c}_{\hbox{{\tiny CNOT-256}}} = 8108$ and the total $T$-count of the SHA-256 oracle $U_g$
(of Fig.~\ref{fig:full_algorithm}) is
\begin{equation}
\add{T^{c}_{U_g} = 2T^c_{\hbox{{\tiny SHA-256}}} + T^{c}_{\hbox{{\tiny CNOT-256}}} = 2\times 228992 + 8108 = 466092}.
\end{equation}
The diffusion operator consists of Clifford gates and a $(k-1)$-fold
controlled-NOT, hence its $T$-count is $T^{c}_{\hbox{{\tiny CNOT-255}}}
= 8076$. The $T$-count of one Grover iteration $G$ is therefore
\begin{equation}
\add{T^{c}_{G} = T^{c}_{U_g} + T^{c}_{\hbox{{\tiny CNOT-255}}} = 466092 + 8076 = 474168},
\end{equation}
and the $T$-count for the full Grover algorithm (\add{let us call it $GA$})
is
\begin{equation}T^c_{GA} = \lfloor\pi/4 \times 2^{128} \rfloor \times 474168\approx
1.27\times 10^{44}.\end{equation}
For this $T$-count the output error rate for state
distillation should be no greater than $p_{out}=1/\add{T^c_{GA}}\approx
7.89\times 10^{-45}$. Assuming a magic state injection
error rate $p_{in}=10^{-4}$, a per-gate error rate
$p_{g}=10^{-5}$, and choosing $\varepsilon = 1$,
Algorithm~\ref{alg:statedistillation} suggests 3 layers of
distillation, with distances $d_1=33, d_2=13$ and $d_3=7$.

The bottom layer of distillation occupies the largest footprint in the
surface code. Three layers of distillation consume $\add{N_{dist}}
= 16 \times 15 \times 15 = 3600$ input states in the process of
generating a single $\ket{A_{L}}$ state. These input states are encoded
on a distance $d_3=7$ code that uses
$2.5 \times 1.25 \times d_3^2 \approx 154$
physical qubits per logical qubit. The total footprint of the
distillation circuit is then
$\add{N_{dist}} \times 154\approx 5.54\times10^5$ physical qubits.
The round of distillation is completed in $10d_3=70$ surface code cycles.

The middle layer of distillation requires a $d_2 = 13$ surface code, for
which a logical qubit takes $2.5 \times 1.25 \times d_2^2 \approx
\rev{530}{529}$ physical qubits. The total number of physical qubits
required in the second round is therefore $16 \times 15 \times
\rev{530}{529}\approx 1.27\times 10^5$ physical qubits, with the round
of distillation completed in $10 d_2 = 130$ surface code cycles.

The top layer of state distillation requires a $d_1 = 33$
surface code, for which a logical qubit takes $2.5 \times 1.25 \times
d_1^2 \approx 3404$ physical qubits. The total number of physical qubits
required in the top layer is therefore $16 \times 3404 = 54464$ physical
qubits, with the round of distillation completed in $10 d_1 = 330$
surface code cycles.

Note that the physical qubits required in the bottom layer of state
distillation can be reused in the middle and top layers. Therefore the
total number of physical qubits required for successfully distilling one
purified $\ket{A_L}$ state is $\add{n_{dist}} = 5.54 \times 10^5$. The
concatenated distillation scheme is performed in $\add{\sigma_{dist} =}
70 + 130 + 330 = 530$ surface code cycles. Since the middle layer of
distillation has smaller footprint than the bottom layer,
distillation can potentially be pipelined to produce
$\add{\phi} = (5.54\times 10^5) /
(1.27\times 10^5) \approx 4$ magic states in parallel.
Assuming, as in
\cite{PhysRevA.86.032324}, a $\add{t_{sc}} = 200$ ns time for a surface
code cycle, a magic state distillery can therefore produce $4$
$\ket{A_L}$ states every $\add{\sigma_{dist} \times t_{sc}} \approx
106\mu$s. Generating the requisite $T^c_{GA} = 1.27\times 10^{44}$ magic states
with a single distillery would take approximately $\add{t_{dist}} =
3.37\times 10^{39}$s $\approx 1.06\times 10^{32}$ years.

We now compute the distance required to embed the entire algorithm in a
surface code. The number of Clifford gates in one iteration of $G$ is
roughly $8.\rev{43}{76}\times 10^6$, so the full attack circuit performs
around $2.\rev{25}{34} \times 10^{45}$
Clifford gates. The overall error rate of the circuit should therefore
be less than $4.\rev{44}{27} \times 10^{-46}$. \rev{}{To compute the
required distance, we seek the smallest $d$ that satisfies the
inequality~\cite{Fowler2012a}}
\begin{equation}
 \left( \frac{p_{in}}{0.0125} \right) ^{\frac{d + 1}{2}} < 4.27 \times 10^{-46},
 \label{eq:fowler_state_dist}
\end{equation}
\noindent \rev{}{and find this to be \add{$d_{\hbox{{\tiny SHA-256}}} =
43$}. The total number of physical qubits in the Grover portion of the
circuit is then $2402 \times
(2.5\times1.25\times43^2) = 1.39 \times 10^7$.}

We can further estimate the number of cycles required to run the entire
algorithm, \add{$\sigma_{GA}$}. Consider a single iteration of $G$ from
Fig.~\ref{fig:full_algorithm}. The $T$-count is \add{$T^{c}_{GA} =
1.27 \times 10^{44}$} and the $T$-depth is \add{$T^{d}_{GA} = 4.79 \times 10^{43}$}
for one iteration of SHA-256, yielding \add{$T^{w}_{G} = T^c_{GA} /
T^d_{GA} \approx 3$}.

Our SHA-256 circuit has \add{$N_{\hbox{{\tiny SHA-256}}}$ = 2402}
logical qubits. Between sequential $T$ gates on any one qubit we will
perform some number of Clifford operations. These are mostly CNOT gates,
which take $2$ surface code cycles, and Hadamard gates, which take a
number of surface code cycles equal to the code distance~\cite{PhysRevA.86.032324}.

Assuming the $8.76 \times 10^6$ Clifford gates in one Grover iteration
are uniformly distributed among the $2402$ logical qubits, then we
expect to perform
$8.76\times 10^6 / (2402 \times T^{d}_{G}) \approx 0.026$
Clifford operations per qubit per layer of depth. As a
magic state distillery produces $4$ magic states per $530$ surface code cycles,
we can perform a single layer of $T$ depth every $530$ surface code
cycles. \add{We thus need only a single distillery, $\Phi =
1$}. On average about $2\%$ of the Cliffords are Hadamards, and the
remaining $98\%$ are CNOTs. This implies that the expected number of
surface code cycles required to implement the $0.025$ average number of
Clifford gates in a given layer of $T$ depth is $2\% \times 0.025 \times
43 + 98\% \times 0.025 \times 2 = 0.071$. As this is significantly lower
than 1, we conclude that performing the $T$ gates comprises the largest
part of the implementation, while the qubits performing the Clifford
gates are idle most of the time. In conclusion, the total number of
cycles is determined solely by magic state production, i.e.
\[\add{\sigma_{GA}} = \lfloor \pi/4\times2^{128}\rfloor \times
530\times (2 T^{d}_{\hbox{{\tiny SHA-256}}} ) \approx 2^{153.8}.\]

As discussed in Sec.~\ref{sct:cost_metric}, the total cost of a quantum
attack against SHA-256 equals the product of the total number of logical
qubits (including the ones used for magic state distillation) and the
number of code cycles, which in our case results in
\[\add{(N_{\hbox{{\tiny SHA-256}}} + \Phi N_{dist}) \sigma_{GA}} = (2402
+ 1 \times 3600)\times 2^{153.8}\approx 2^{166.4},\]\rev{}{corresponding
to an overhead factor of $v = (166.4 - 128)/\log_2(256) = 4.8$}.

\subsection{SHA3-256}\label{sct5B}

We perform a similar analysis for SHA3-256. \add{We have $T^{c}_{U_g} = 2\times499200 + 32\times256 - 84 = 1006508$, and $T^{c}_G = 1006508 + 32\times 255 - 84 = 1014584$, and thus the full Grover algorithm takes $T$-count $T^c_{GA} = \lfloor \pi/4 \times 2^{128} \rfloor \times 1014584 \approx 2.\rev{70}{71} \times 10^{44}$}. If we choose, like in the case of SHA-256, $p_{in} = 10^{-4}, p_g = 10^{-5}$, and $\varepsilon = 1$, Algorithm \ref{alg:statedistillation} yields 3 layers of distillation with distances $d_1 = 33, d_2 = 13$, and $d_3 = 7$; these are identical to those of SHA-256. Thus, the distillation code requires takes 3600 logical qubits (and $5.54\times 10^5$ physical qubits), and in 530 cycles is able to produce roughly 4 magic states.

We compute the distance required to embed the entire algorithm in a surface code. The total number of Cliffords in one iteration of $G$ is roughly $6.\rev{89}{90}\times 10^7$, so the total number will be around $1.84 \times 10^{46}$ operations. We thus need the overall error rate to be less than $5.43 \times 10^{-47}$, which by Eq. \ref{eq:fowler_state_dist} yields a distance $d_{\hbox{{\tiny SHA3-256}}} = \rev{43}{44}$. \rev{}{The number of physical qubits is then $1.94 \times 10^7$.}

Consider a single iteration of $G$ from Fig.~\ref{fig:full_algorithm}. $T^c_G = 1014584$ and $T^d_{\hbox{{\tiny SHA3-256}}} = 432$, which yields $T^w_G = 1014584 / (2\times432) = 1175$.
Above we figured we can compute 4 magic states in 530 code cycles. Then, to compute 1175 magic states in the same number of cycles we will need roughly \add{$\Phi = 294$} distillation factories working in parallel to keep up. This will increase the number of physical qubits required for state distillation to $1.63\times 10^8$. If we assume \add{$t_{sc} = 200$}ns cycle time, generation of the full set of magic states will take $2.\rev{44}{28}\times 10^{37}$s, or about $\add{t_{dist}} = 7.\rev{72}{23} \times 10^{29}$ years.

Our SHA3-256 circuit uses \add{$N_{\hbox{{\tiny SHA3-256}}}$ = 3200} logical qubits.
Assuming the $6.90\times10^7$ Clifford gates per Grover iteration are
uniformly distributed among the qubits, and between the $864$ sequential
$T$ gates, we must be able to implement $6.\rev{89}{90}\times 10^7 /
(3200 \times 864) \approx 25$ Clifford operations per qubit per layer of $T$-depth. As the
ratio of CNOTs to Hadamards is roughly $202$ to $1$, i.e. 99.5\% of the
Cliffords are CNOTs and only $0.5\%$ are Hadamards, the expected number
of surface code cycles required to implement the \add{average of }$25$
Clifford gates in a given layer of $T$ depth is $25\times (0.005 \times
\rev{43}{44} + 0.995 \times 2) \approx 55$. We have used just
enough ancilla factories to implement a single layer of $T$-depth in $530$
cycles, meaning that once again the limiting step in implementing this
circuit is the production of magic states. Hence, we can compute the
total number of surface code cycles required to implement SHA3-256 using
just the $T$-depth: \[\add{\sigma_{GA} =} \lfloor \pi/4 \times 2^{128} \rfloor \times 530
\times (2 \add{T^{d}_{\hbox{{\tiny SHA3-256}}}}) \approx 1.22 \times
10^{44} \approx 2^{146.5}.\]

\break
The total cost of a quantum attack agains\rev{}{t} SHA3-256 \add{is
then}
\[\add{(N_{\hbox{{\tiny SHA3-256}}} + \Phi N_{dist}) \sigma_{GA}}
= (3200 + 294 \times 3600)\times 2^{146.5}\approx 2^{166.5},\]\rev{}{or an overhead of $v = (166.5 - 128)/\log_2(256) = 4.81$}.

% Vlad
\begin{table}
\footnotesize
\centering
\renewcommand*{\arraystretch}{1.2}
\begin{tabular}{c | l r c r}
  \hline\hline
  && \textbf{SHA-256} && \textbf{SHA3-256} \\ \hline
\parbox[t]{4mm}{\multirow{5}{*}{\rotatebox[origin=c]{90}{Grover}}}
  & $T$-count & $1.27 \times 10^{44} $  &\quad\quad & $2.71 \times 10^{44}  $ \\
  & $T$-depth & $3.76 \times 10^{43}$ & &$2.31 \times 10^{41} $ \\
  & Logical qubits & $2402$ & &$3200$ \\
  & Surface code distance & 43 && \rev{43}{44} \\
  & Physical qubits & $1.39 \times 10^7$ && $1.\rev{85}{94} \times 10^7$  \\ \hline
\parbox[t]{4mm}{\multirow{5}{*}{\rotatebox[origin=l]{90}{~Distilleries}}}
  & Logical qubits per distillery & $3600$ &&  $3600$ \\
  & Number of distilleries & 1 && 294 \\
  &Surface code distances & $\{33, 13, 7\}$ && $\{33, 13, 7\}$ \\
  &Physical qubits &  $5.54\times 10^5$ && $1.63 \times 10^8$ \\ \hline
\parbox[t]{4mm}{\multirow{3}{*}{\rotatebox[origin=c]{90}{Total}}}
  &Logical qubits & $2^{12.6} $ && $2^{20} $ \\
  &Surface code cycles & $2^{153.8} $ && $2^{146.5} $ \\
  &Total cost & $2^{166.4}$ && $2^{166.5}$ \\ \hline \hline
\end{tabular}
\caption{Fault-tolerant resource counts \add{for Grover search of} SHA-256 and SHA3-256.}
\label{tab:faulttolsummary}
\end{table}
\section{Conclusions and open questions\label{sct:conclusions}}
We estimated the cost of a quantum pre-image attack on SHA-256 and SHA3-256 cryptographic hash functions via Grover's quantum searching algorithm. We constructed reversible implementations of both SHA-256 and SHA3-256 cryptographic hash functions, for which we optimized their corresponding $T$-count \rev{}{ and depth}. We then estimated the required physical resources needed to run a brute force Grover search on a fault-tolerant surface code based architecture.

We showed that attacking SHA-256 requires approximately $2^{153.8}$ surface code cycles and that attacking SHA3-256 requires approximately $2^{146.5}$ surface code cycles\rev{, respectively}{}. For both SHA-256 and SHA3-256 we found that the total cost when including the classical processing \rev{raises}{increases} to approximately $2^{166}$ basic operations.%, hence (with our assumptions) both SHA hash functions provide around 162 bits of security against a quantum adversary.

Our estimates are by no means a lower bound, as they are based on a series of assumptions. First, we optimized our $T$-count by optimizing each \rev{individual components}{component} of the SHA oracle \rev{}{individually}, which of course is not optimal. Dedicated optimization schemes may achieve better results. Second, we considered a surface code fault-tolerant implementation, as such a scheme looks the most promising \rev{today}{at present}. However it may be the case that other quantum error correcting schemes \rev{may}{} perform better. Finally, we considered an optimistic per-gate error rate of about $10^{-5}$, which is the limit of current quantum hardware. This number will probably be improved in the future. Improving any of the issues listed above will certainly result in a better estimate and a lower number of operations, however the decrease \rev{}{in the number of bits of security} will likely be limited.

\section*{Acknowledgments}
We acknowledge support from NSERC and CIFAR.
IQC and PI are supported in part by the Government of Canada and the Province of Ontario.

\bibliographystyle{splncs04} % Modified the splncs03 to make the references come in numerical order
%\bibliography{breaking_sha.bib}

% Olivia
\appendix
%\begin{titlepage}
%\setcounter{page}{21}
%\section{Flowchart for analyzing Grover's algorithm\label{apdxA}}
%\begin{figure*}[ht!]
%  \includegraphics[scale = 0.58]{images/flowchart_full.eps}
%  \caption{Flowchart for computation of resources to perform a pre-image attack on a hash function $f : \{0, 1\}^k \rightarrow \{0, 1 \}^k$.}
%  \label{fig:flowchart_full}
%  \vspace{2cm}
%\end{figure*}
%\end{titlepage}

% John
%\input{appendix_B/appendix_b.tex}
\section{Tables}\label{app:tables}
%   \begin{table}[h!]
%     \begin{tabular}{cc|cccccccc}
%     \multicolumn{2}{c|}{\multirow{2}{*}{k-C}} &\multicolumn{8}{c}{$C$}\\
%        && 16 & 32 & 48 & 64 & 80 & 96 & 112 & 128\\\hline
%   \multirow{6}{*}{$a$}
%   & $0$ & 16 & 32 & 48 & 64 & 80 & 96 & 112 & 128\\
%   & $1$ & 6 & 20 & 35 & 50 & 65 & 81 & 96 & 112\\
%   & $2$ & 0 & 10 & 23 & 37 & 51 & 66 & 81 & 96\\
%   & $3$ & -5 & 1 & 12 & 25 & 38 & 52 & 67 & 81\\
%   & $4$ & -9 & -6 & 2 & 13 & 26 & 39 & 53 & 67\\
%   & $5$ &-10 & -12 & -6 & 3 & 14 & 26 & 39 & 53
%     \end{tabular}
%     \caption{The advantage, $k-C$, of a quantum pre-image search that can be performed for
%       cost $2^C = k^a2^{k/2}$. The negative entries correspond to a regime where quantum search is strictly worse than classical search.}
%     \label{tab:kCtov}
%   \end{table}

\begin{table}[H]
 \centering
  \begin{tabular}{cc|cccccccc}
    \multicolumn{2}{c|}{\multirow{2}{*}{ $\frac{k(a,C)}{C}$ }} &\multicolumn{8}{c}{$C$}\\
     && 16 & 32 & 48 & 64 & 80 & 96 & 112 & 128\\\hline
\multirow{6}{*}{$a$}
& $0$ & 2.00 & 2.00 & 2.00 & 2.00 & 2.00 & 2.00 & 2.00 & 2.00\\
& $1$ & 1.38 & 1.63 & 1.73 & 1.78 & 1.81 & 1.84 & 1.86 & 1.88\\
& $2$ & 1.00 & 1.31 & 1.48 & 1.58 & 1.64 & 1.69 & 1.72 & 1.75\\
& $3$ & 0.69 & 1.03 & 1.25 & 1.39 & 1.48 & 1.54 & 1.60 & 1.63\\
& $4$ & 0.44 & 0.81 & 1.04 & 1.20 & 1.33 & 1.41 & 1.47 & 1.52\\
& $5$ & 0.38 & 0.63 & 0.88 & 1.05 & 1.18 & 1.27 & 1.35 & 1.41
  \end{tabular}
  \caption{The advantage, $k/C$, of a quantum pre-image search that can be performed for
    cost $2^C = k^a2^{k/2}$. The entries less than $1$ correspond to a regime where quantum search is strictly worse than classical search.}
  \label{tab:kCtov}
\end{table}

%
%   \begin{table}[h!]
%     \begin{tabular}{cc|cccccccc}
%   \multicolumn{2}{c|}{\multirow{2}{*}{C}} &\multicolumn{8}{c}{$k$}\\
%        && 32 & 64 & 96 & 128 & 160 & 192 & 224 & 256\\\hline
%   \multirow{6}{*}{$a$}
%   & $0$ & 16 & 32 & 48 & 64 & 80 & 96 & 112 & 128\\
%   & $1$ & 21 & 38 & 55 & 71 & 87 & 104 & 120 & 136\\
%   & $2$ & 26 & 44 & 61 & 78 & 95 & 111 & 128 & 144\\
%   & $3$ & 31 & 50 & 68 & 85 & 102 & 119 & 135 & 152\\
%   & $4$ & 36 & 56 & 74 & 92 & 109 & 126 & 143 & 160\\
%   & $5$ & 41 & 62 & 81 & 99 & 117 & 134 & 151 & 168
%     \end{tabular}
%     \caption{The log cost $C = \log_2{k^a2^{k/2}}$ of $k$-bit search with overhead $a$.}
%     \label{tab:katoC}
%   \end{table}

\begin{table}[H]
 \centering
\begin{tabular}{c|ccccc}
$a$ & 1 & 2  & 3  & 4  & 5 \\\hline
$k$ & 5 & 16 & 30 & 44 & 59
  \end{tabular}
  \caption{The $k$ for which the classical and quantum search costs are equal after accounting
  for the $k^a$ overhead for quantum search.}
  \label{tab:crossover}
\end{table}

\section{Parallel quantum search}

%\subsection{Parallel quantum search}

Classical search is easily parallelized by distributing the $2^k$ bitstrings
among $2^t$ processors. Each processor fixes the first $t$ bits of its input to
a unique string and sequentially evaluates every setting of the remaining $k-t$
bits. Since our cost metric counts only the number of invocations of $g$, the
cost of parallel classical search is $2^k$ for all $t$. If one is more
concerned with time (i.e.  the number of sequential invocations) than with
area, or vice versa, it may be more useful to report the cost as $(T, A)$. Or,
in this case, $(2^{k-t}, 2^t)$.

Quantum computation has a different time/area trade-off curve. In
particular, parallel quantum strategies have strictly greater cost than
sequential quantum search. Consider sequential quantum search with cost
$C(1) = (C_T, C_A) = (k^a 2^{k/2}, k^b)$.  Parallelizing this algorithm
across $2^t$ quantum processors reduces the temporal cost per processor
by a factor of $2^{t/2}$ and increases the area by a factor of $2^t$.
Fixing $t$ bits of the input does not change the overhead of the Grover
iteration, so the cost for parallel quantum search on $2^t$ processors
is $C(2^t) = (2^{-t/2} C_T, 2^{t} C_A) = (k^a2^{(k-t)/2}, k^b2^t)$.

% \subsection{Multi pre-image search}
% The setup is the same as pre-image search, but we are given a collection of $M$ target values
% $Y = \{y_i : i \in [M]\}$ and are asked to find $x$ such that $f(x) \in Y$. Classical search
% will find one such value in $2^n/M$ queries, and will find \emph{all} such values in $2^k$
% queries. Grover's algorithm on a single quantum processor will take $\sqrt{2^k/M}$ \cites{}.

\end{document}